\documentclass[nofootinbib]{revtex4}
\usepackage{amsfonts,amsmath}
\usepackage{graphicx}

\newtheorem{theorem}{Theorem}
\newtheorem{lemma}{Lemma}
\newcommand{\qed}{\nobreak \ifvmode \relax \else
      \ifdim\lastskip<1.5em \hskip-\lastskip
      \hskip1.5em plus0em minus0.5em \fi \nobreak
      \vrule height0.75em width0.5em depth0.25em\fi}

\def\<{\langle}
\def\>{\rangle}
\def\l{\left}
\def\r{\right}
\newcommand{\ket}[1]{|#1\rangle}
\newcommand{\bra}[1]{\langle #1|}

\newcommand{\proj}[1]{\left|#1\right\>\!\left\<#1\right|}
\newcommand{\ot}{\otimes}
\newcommand{\eps}{\epsilon}

\def\benum{\begin{enumerate}}
\def\eenum{\end{enumerate}}
\def\bit{\begin{itemize}}
\def\eit{\end{itemize}}

\newcommand{\be}{\begin{equation}}
\newcommand{\ee}{\end{equation}}
\newcommand{\tr}{\mathop{\mathrm{tr}}\nolimits}
\def\ba#1\ea{\begin{align}#1\end{align}}
\def\ban#1\ean{\begin{align*}#1\end{align*}}

\newcommand{\eq}[1]{(\ref{eq:#1})}

\newcommand{\secref}[1]{Section~\ref{sec:#1}}
\newcommand{\lemref}[1]{Lemma~\ref{lem:#1}}
\newcommand{\thm}[1]{Theorem~\ref{thm:#1}}

\DeclareMathOperator{\poly}{poly}

\DeclareMathOperator{\Span}{Span}
\DeclareMathOperator{\Vol}{Vol}

\def\bP{{\mathbf{P}}}

\def\bbC{{\mathbb{C}}}
\def\bbE{{\mathbb{E}}}

\def\cE{{\cal E}}
\def\cO{{\cal O}}

\def\cS{{\cal S}}
\def\cT{{\cal T}}
\def\cU{{\cal U}}

\def\d{{\rm d}}

\begin{document}
\title{Classical and Quantum Tensor Product Expanders}
\author{M.~B.~Hastings}
\affiliation{Center for Nonlinear Studies and Theoretical Division,
Los Alamos National Laboratory, Los Alamos, NM, 87545}
\author{A.~W.~Harrow}
\affiliation{Department of Computer Science, University of Bristol,
  Bristol, U.K.}
\begin{abstract}
We introduce the concept of quantum tensor product expanders.  These
generalize the concept of quantum expanders, which are quantum maps that
are efficient randomizers and use only a small number of Kraus operator.
Quantum tensor product expanders
act on several copies of a given system, where
the Kraus operators are tensor products of the Kraus operator
on a single system.  We begin with the classical case, and show that
a classical two-copy expander can be used to produce a quantum expander.
We then discuss the quantum case and give applications to the Solovay-Kitaev
problem.
We give probabilistic constructions in both classical and quantum
cases, giving tight bounds on the expectation value of the
largest nontrivial eigenvalue in the quantum case.
\end{abstract}
\maketitle

\section{Background: classical and quantum expanders}
\subsection{Definitions}
The concept of $t$-designs\cite{tdesigns} provides a way of
randomizing quantum states.  For example, a $1$-design is a set of
unitaries $\{U_k\}$, where $k=1,\ldots,K$, such that the average over
the set takes any input state to a maximally mixed state.  A
$2$-design is a set of unitaries such that applying $U_k\otimes U_k$
to a state on a bipartite system generates the twirling
operation\cite{twirl}.  Quantum expanders, as studied
in Hamiltonian complexity\cite{entqgs}, computer
science\cite{qexp}, and quantum information theory\cite{qexp2},
provide a way of approximately realizing a $1$-design by repeatedly
applying a completely positive map built out of a small number of
unitaries.  In this paper, we introduce the concept of ``tensor
product expanders'', which generalize this result and give us a way to
approximately realize $t$-designs.  We also discuss the classical
case, and show that classical tensor product expanders can be used to
generate quantum expanders.

Quantum expanders are a quantum analogue of expander graphs\cite{expanderbook}.
In the quantum case, we consider a completely positive, trace preserving map
\be
\label{mapdefn}
{\cal E}(M)=\sum_{s=1}^D A^{\dagger}(s) M A(s),
\ee
where the number of Kraus operators $D$ is relatively small and the
map ${\cal E}$ has a spectral gap between the largest eigenvalue,
equal to unity, and the next largest eigenvalue\footnote{In the non-Hermitian case discussed below, we define the gap instead to be one minus the second-largest singular value of the map ${\cal E}$.}.  We write the spectrum of
${\cal E}$ as $\lambda_1,\lambda_2,...$ with $\lambda_1=1$ and
$\lambda_2,...$ all bounded in absolute value by some $\lambda<1$.
We can equivalently consider the operator $\hat{\cE}:=\sum_{s=1}^D
A(s) \ot A(s)^*$.

In
this paper we consider the case in which the operators
$A^{\dagger}(s)$ are proportional to unitary operators:
\be A(s)=\frac{1}{\sqrt{D}}U(s).  \ee
Then the expander map can be implemented by choosing $s$ uniformly at
random from $\{1,\ldots,D\}$, and then applying $U(s)$ to the quantum
state.  The natural generalization of this process, in which we
consider $k$ copies of a quantum system, choose a unitary at random,
and apply the unitary to all $k$ copies, will be called a $k$-copy
tensor product expander.  We will show that these give a way to
approximate $t$-designs for $t=k$.

Random walks on expander graphs can be viewed similarly, as acting
on a distribution with a randomly chosen permutation matrix.
Consider a directed
graph, where each node has $D$ edges leaving it.
Label the edges from $1$ up to
$D$ such that each label appears exactly once among the incoming edges
of each vertex and exactly once among the outgoing edges of each vertex.  Then, for each edge label $s$, $1\leq s\leq D$,
define a permutation $\pi_s$, where $\pi_s(i)=j$ if a directed edge
with label $s$ goes from node $i$ to node $j$.  Then, given a random walk
on the graph, the probability distribution $p(i)$ changes in a single step
by
\be
p(i) \rightarrow \frac{1}{D} \sum_{s=1}^D \sum_{j=1}^N P(s)_{ij} p(j),
\ee
where $P(s)$ is the permutation matrix corresponding to the permutation
$\pi_s$; i.e. $P(s)_{ij}=1$ if $\pi_s(j)=i$ and 0 otherwise.

{\em Hermitian expanders:} It is sometimes convenient to guarantee
that an expander we construct is Hermitian.  To obtain Hermitian
${\cal E}$ in the quantum case, we impose
\be
\label{impose}
U(s+D/2)=U(s)^{\dagger}.
\ee
Similarly, in the classical case, we impose
\be
\pi_s=\pi_{s+D/2}^{-1}
\ee
This turns the directed graph into an undirected graph.
For notational
convenience, we identify $s+D$ with $s$ throughout this paper, so that
$s$ is a periodic variable with period $D$.  Note that this constraint
(\ref{impose}) requires that $D$ be even.  There do exist other
ways to construct Hermitian expanders
with odd $D$, if for some $s$ we have $U(s)=U(s)^{\dagger}$.

\subsection{Application to state randomization}
For classical expanders, an important implication of the spectral gap
is that random walks on an expander graph rapidly approach the
stationary distribution.  Similarly, quantum expanders can be shown to be rapid mixing.  This has application to the problem of {\em state randomization}, in which classical randomness is used to map a quantum state to an output that is close in trace distance to the maximally mixed state.  Ideally the constructions would be [computationally] efficient, meaning they run in time polynomial in the number of qubits, and would use as few random bits as possible.

To make this concrete, suppose that ${\cal E}$ is Hermitian and unital with gap $1-\lambda$,
  and consider
a quantum state $\rho$.  We wish to bound the trace norm distance between the
maximally mixed state and the state
${\cal E}^m(\rho)$ obtained by acting on $\rho$ with some high
power, $m$, of the map ${\cal E}$.
 The calculation exactly
follows the classical case.  We begin by bounding the
$\ell_2$
distance.  For a matrix $A$, define $\|A\|_2=\sqrt{\tr A^\dag A}$ and
$\|A\|_1=\tr |A| = \tr\sqrt{A^\dag A}$.  Then
\be
\label{eq:l2mix}
\l\| \cE^m(\rho) - \frac{\openone}{N}\r\|_2^2\leq
|\lambda|^{2m},
\ee
as may be shown by writing $\rho$ as a linear combination of
eigenvectors of ${\cal E}$,
and then by Cauchy-Schwartz,
\be
\l\|{\cal E}^m(\rho)-\frac{\openone}{N}\r\|_1\leq
\sqrt{N} |\lambda|^{m}.
\ee
Thus, to obtain a given bound on the trace norm distance $\epsilon$,
it suffices to take
\be
m\geq \log_{\lambda}(\epsilon/\sqrt{N}).
\label{eq:1-d-num-iters}
\ee
This implies that the set of unitaries, consisting of all unitaries of
the form $U(s_1)U(s_2)\cdots U(s_m)$, gives an $\eps$-approximate
$1$-design using 
\be K := D^m = \l(\frac{N}{\eps^2}\r)^{\frac{1}{2}\log_{1/\lambda}(D)}
\label{eq:1-d-num-gates}\ee
unitaries.  

The exponent $\frac{1}{2}\log(D)/\log(1/\lambda)$ can be thought of
as a measure of the efficiency of an expander, meaning the number of
bits of randomness it requires to achieve a certain amount of state
randomization.  Before showing how to evaluate
$\frac{1}{2}\log(D)/\log(1/\lambda)$, we review other methods of
$\ell_1$ state randomization.  The simplest is to apply one of $N^2$
generalized Pauli operators.  This can be done efficiently (i.e. in
time $\poly\log(N)$) and
perfectly randomizes any state (i.e. $\eps=0$).  However, it uses far
more randomness than necessary when $\eps>0$.  Choosing
$K=O(N\eps^{-2}\log(1/\eps))$ random unitaries was shown to suffice in
\cite{aubrun}, improving a result of
\cite{HLSW-random} (both of which in fact addressed the more difficult
problem of $\ell_\infty$ state randomization).  Similarly an efficient
$K=4N\eps^{-2}$ construction was given in
\cite{DN-random}, which uses less randomness than the efficient constructions of
\cite{AS-random} and even than the inefficient constructions based on
random unitaries.  We note in passing that the constructions in
\cite{DN-random,AS-random} are based on expanders with
$\lambda=\eps/\sqrt{N}$ and $D=K$.

An expander-based state randomization scheme will be efficient if the
underlying expander is efficient and the number of unitaries it uses
will be given by \eq{1-d-num-gates}.  Unfortunately
$\frac{1}{2}\log(D)/\log(1/\lambda)$ is larger than 2 for all known
efficient constant-degree expander
constructions\cite{qexp2,qexp3,qexp4} (e.g. for the Margulis
expander\cite{qexp3}, it is $\approx 8.4$, and for the zig-zag
product\cite{qexp2} 
it is $2+o(1)$).  However, if 
$U(1),\ldots,U(D/2)$ are chosen at random with $U(s+D/2)=U(s)^{\dagger}$
then Ref.~\cite{sd} showed that
with high probability
$\frac{1}{2}\log(D)/\log(1/\lambda) \approx 1+\cO(\log(N) N^{-1/6})+2/\log(D)$,
and thus that $K$
is within a small multiplicative factor of $N/\eps^2$.

We summarize the above discussion as follows:
\begin{theorem}\label{thm:state-random}
For any $N$ and any $\eps>0$, consider a set of unitaries
$U_1,\ldots,U_K\in\cU_N$, which are
taken to be strings of
unitaries drawn from a set of $D/2$ unitaries $U(1),...,U(D/2)$ and their
conjugates for any $D\geq 4$.
Then for most choices of $U(1),...,U(D/2)$,  choose the
string length such that \be K=\Bigl(\frac{N}{\eps^2}\Bigr)^{1+O(N^{-1/6}\log(N))+2/\log(D)}\ee and
$$\l\|\frac{1}{K}\sum_{s=1}^K U_s\rho U_s^\dag -
\frac{\openone}{N}\r\|_1 \leq \eps,$$
for all $N$-dimensional density matrices $\rho$.
\end{theorem}

If we take $D\approx 4N/\eps^2$ then \thm{state-random} can be thought
of as tightening the analysis of random unitaries from
\cite{aubrun,HLSW-random,DN-random}, so that only $(4+o(1))N/\eps^2$
random unitaries are necessary.  This shows that Haar-uniform
unitaries require almost exactly the same amount of randomness as the
construction of \cite{DN-random}, although they have the substantial
disadvantage of requiring $\poly(N)$ time to implement instead of
$\poly(\log(N))$ time.  Since $\lambda \geq (2\sqrt{D-1}/D - O(1/N))\cdot(1-
O(\log\log(N) / \log(N)))$ for any
quantum expander that includes its own inverses
\cite{sd}, one can show that $4N/\eps^2$ is the minimum possible
values of $K$ for any expander-based randomizing map.

Apart from random unitaries and the large-$D$ constructions of
\cite{AS-random,DN-random}, we know of one other class of quantum expanders
for which 
$\frac{1}{2}\log(D)/\log(1/\lambda)\approx 1$.  These are obtained by
applying the prescription of \cite{qexp4} to the $SU(2)$ expanders
described by Lubotsky, Phillips and Sarnak in \cite{LPS}.  Such expanders exist for any $N$ whenever $D$ is odd and
$2D-1$ is prime, and satisfy $\lambda=2\sqrt{D-1}$ exactly.  Thus,
they provide another $K\approx 4N/\eps^2$ method of performing state
randomization.  However, the only claimed efficient construction of
these expanders\cite{Zalka} has an incomplete proof.

In the non-Hermitian case, \eq{l2mix} holds when $\lambda$ is the second-largest singular value of an expander.  If $U(1),\ldots,U(D)$ are chosen uniformly at random, then \cite{sd} proved that with high probability the singular values of ${\cal E}^m$ for $m=\cO(N^{1/6})$
are bounded by $N^2(1/\sqrt{D})^m(1+o(1))$.  This implies that the second-largest {\em eigenvalue} of $\cE$ is $\leq
\frac{1}{\sqrt{D}}(1+O(\log(N)N^{-1/6}))$, but does not yield meaningful bounds on the second-largest singular value of $\cE$.  Indeed, Tobias Osborne has pointed out that when $m=1$ and $D=2$, the second largest singular value is equal to unity.   If ${\cal
E}^m$ turned out to have singular values nearly equal to $D^{-m/2}$
then it would imply that $\approx N/\eps^2$ random unitaries sufficed
to $\eps$-randomize a state.

We now turn to tensor product expanders, considering classical tensor
product expanders in \secref{classical-TPE} and quantum tensor product
expanders in \secref{quantum-TPE}.  The mixing analysis above generalizes
in the tensor product case to give approximate $t$-designs.  We will describe
randomized constructions of both classical and quantum tensor product
expanders.  Our basic tool to prove that a random construction gives
an expander with high probability is the trace method (see, for
example \cite{expanderbook,bs}).  The basic idea of the trace method is
to bound eigenvalues of some linear operator by bounding the trace of
high powers of that operator.  
For example, for a positive definite Hermitian operator whose two
largest eigenvalues are equal to unity and to $\lambda$, the trace of
the $m^{\text{th}}$ power is at least equal to $1+\lambda^m$, so by bounding the
trace we bound $\lambda$.
We focus on high powers of the operator
so that the trace will be dominated by the largest eigenvalues.  The
trace method will be adapted, with slight modifications, to the various
cases, depending on whether classical or classical and quantum, and
depending on whether we consider an expander and or a tensor product expander.

\section{Classical Tensor Product Expanders}\label{sec:classical-TPE}

In this section we define classical tensor product expanders, and give
a random construction of them.  We then show an application of them to
constructing quantum expanders.

\subsection{Preliminaries, Definitions and Applications }
We define an $(N,D,\lambda,k)$ classical $k$-copy tensor product expander
to be a set of $N$-by-$N$ permutation matrices $P(s)$, $1\leq s \leq D$,
with the property that the matrix $L$, defined by
\be
\label{Ldef}
L_k=\frac{1}{D} \sum_{s=1}^D P(s)^{\otimes k}
\ee
has
some number, $f_k^N$, eigenvalues equal to unity, with $f_k^N$ defined below,
and then all other eigenvalues less than
or equal to $\lambda$ in absolute value.  (Again, if $L_k$ is non-Hermitian then we consider its singular values.)

We can obtain Hermitian operators $L_k$ by considering $D$ even,
and imposing $P(s+D/2)=P(s)^{\dagger}$.  
To obtain Hermitian $L_k$
for $D$ odd, we can instead
impose $P(s)=P(s)^{\dagger}$; that is, the permutation
matrices correspond to perfect matchings.
Both models corresponds to models
of random graphs for $k=1$ discussed in \cite{friedman}.  

These expanders can also be defined by
graphs with $N^k$ nodes, labelled $(n_1,n_2,\ldots,n_k)$,
where $1\leq n_i\leq N$.  There is an edge from one node $(n_1,\ldots,n_k)$ 
to another node $(n_1',\ldots,n_k')$ if and only if one of the given
permutations sends $n_1\rightarrow n_1',\ldots,n_k\rightarrow n_k'$.
We refer to this graph as $G_k$.
Alternatively, we can
regard $n_1,\ldots,n_k$ as $k$ different random walkers executing a correlated
random walk on the original graph.

The function $f_k^N$ is defined to be equal to the number of unit eigenvalues
of the operator
\be
\frac{1}{N!} \sum_{\pi\in\cS_N} P_\pi^{\ot k}
\label{eq:ideal-cl-TPE}
\ee
where the sum ranges over {\it all} permutations $\pi$, and $P_\pi$ is the
permutation matrix corresponding to permutation $\pi$.  Since this operator performs an average over a group action, it is a projector.  	Applying it to a computational basis state $\ket{n_1,\ldots,n_k}$ maps it to the superposition of all $\ket{n_1',\ldots,n_k'}$ such that $n_i'=n_j'$ iff $n_i=n_j$.  Thus we can represent eigenstates by partitions of $\{1,\ldots,k\}$ into $\leq N$ blocks, such that indices are equal within blocks and unequal across blocks.  For example, $f_1^N=1$,
 $f_2^N=2$ 
(corresponding to the sum of all states with $n_1=n_2$ and the sum of all states
with $n_1\neq n_2)$, $f_3^N=5$ (corresponding to the possibilities
$n_1=n_2=n_3$, $n_1=n_2\neq n_3$, $n_1=n_3\neq n_2$, $n_2=n_3\neq n_1$, and $n_1\neq n_2
\neq n_3 \neq n_1$), and so on.   Note that if $N\geq k$ then the constraint that there be $\leq N$ blocks becomes superfluous, and $f_k^N$ becomes simply the
$k^{\text{th}}$ Bell number $B_k$, which counts the total number of ways of
partitioning a $k$-element set.
 
Any matrix $L_k$ of the form (\ref{Ldef}) is block diagonal with
$f_k^N$ different blocks depending on the symmetry of the elements
$n_1,\ldots,n_k$ under permutation; we call these subspaces
$S_1,S_2,\ldots,S_{f_k^N}$.  By the arguments of the above paragraph, we
can write the projector in \eq{ideal-cl-TPE} as 
$$\sum_{a=1}^{f_k^n}\proj{u_a},$$
for some unit vectors $\ket{u_a}\in S_a$.  These $\ket{u_a}$ are unit
eigenvalues not only of \eq{ideal-cl-TPE} but also any $L_k$. 

{\em Rapid mixing:} Given the spectral gap, repeatedly applying a
classical tensor product expander many times (of order $k\log(N)$)
generates an approximately $k$-wise independent permutation.  This
means that the results of applying it to $k$ distinct elements are
almost indistinguishable from applying a single permutation to each of the $k$
elements.  More precisely, given an initial probability distribution,
$p$, in any of the $f_k^N$ different subspaces $S_{a}$, we have \be
\Vert L_k^mp-u_a \Vert_1 \leq \sqrt{N}^k |\lambda|^m, \ee where $u_a$ is
the $l_1$ normalized eigenvector with eigenvalue unity in this
subspace.  This approach towards generating $k$-wise independent
permutations has also been considered in \cite{ctpe}.

{\em Expanders are not always tensor product expanders.}  The
requirement that a set of permutations form a tensor product expander
for $k>1$ copies is more stringent than the requirement for $k=1$
copy, as it implies that the correlations between elements are
destroyed by the expander.  For an example of a classical expander
that does not give a tensor product expander, consider any set of $D$
permutation matrices, $P(s)$, on $N$ elements that gives a classical
expander.  Define a new set of permutation matrices, $P'(s)$, on $2N$
elements, such that $P'(s)=P(s)\oplus P(s)$ for $s=1,\ldots,D$.
Finally, define the permutation $P'(D+1)$ which sends $i$ to $i+N$ if
$i\leq N$, and sends $i$ to $i-N$ if $i>N$.  Then, these $D+1$
different permutation matrices define a $k=1$ expander (they simply
correspond to two copies of the original graph, with the possibility
of moving between the two copies by using permutation matrix
$P'(D+1)$), but does not define a $k=2$ expander: if two walkers,
$n_1,n_2$ originally are in the same copy as each other, then they
remain in the same copy.

Another example comes from Cayley graphs.  If $G$ is a group with
generators $g_1,\ldots,g_D$ then the Cayley graph on $G$ is defined by
taking $N=|G|$ and $P(s)\ket{g} = \ket{g_sg}$ for $s=1,\ldots,D$.
There are many Cayley graph expanders known (c.f. Section 11 of
\cite{expanderbook}), but applying $P(s)\ot P(s)$ to any $\ket{g}\ot
\ket{h}$ produces a new state $\ket{\tilde g}\ot\ket{\tilde h}$ with
$\tilde g^{-1}\tilde h=g^{-1}h$.  Thus, no Cayley graph expander
can be a tensor product expander unless it is modified in some way.

{\em The limit of large $k$:}  Observe that any $k$-copy tensor
product expander is also a $k'$-copy tensor product expander for all
$k'\leq k$.  On the other hand, even if $k>N$ then the $k$ walkers can
still occupy only at most $N$ positions.  Thus if a map is an $N$-copy
tensor product expander than it is also a $k$-copy tensor product
expander for all $k$.

An equivalent condition to $\{\pi_1,\ldots,\pi_D\}\subset\cS_N$ being
an $N$-tensor product expander is that the Cayley graph generated by
$\{\pi_1,\ldots,\pi_D\}$ is an expander.  The spectrum of this Cayley
graph is identical (up to multiplicity) to that of $L_k$ for all
$k\geq N$ (with $P(s)$ defined to be $P_{\pi_s}$).

\subsection{Random permutations are tensor product expanders}

The question then naturally arises whether $k>1$ tensor product expanders
actually exist.  Of course there is a trivial $D=N!$ construction
where we take $\{\pi_1,\ldots,\pi_N\}=\cS_N$ and achieve $\lambda=0$
for all $k$.  We would prefer, though, that $D=O(1)$.
The construction of \cite{ctpe} nearly achieves this with
$D=\poly\log(N)$ and $\lambda = 1-1/\poly(k,\log N)$.  For a constant
degree construction, we can use Kassabov's expander\cite{Kass} on $\cS_N$.  This
achieves $D=O(1)$ and $\lambda$ equal to a constant strictly smaller
than 1 for all $N$ and $k$.  Additionally, it can be implemented in
time $\poly\log(N)$.

In this section, we give a randomized construction of tensor product
expanders for any even $D\geq 4$ and with $\lambda \approx
\lambda_H^{\frac{1}{k+1}}$, where 
\be\label{lHdef}\lambda_{H}:= \frac{2\sqrt{D-1}}{D}.\ee
\begin{theorem}\label{thm:random-cl-TPE}
Choose $\pi_1,\ldots,\pi_{D/2}\in\cS_N$ at random and then take
$\pi_{s+D/2}=\pi_s^{-1}$.  Let $P(s)=P_{\pi_s}$.
For any $k$, let $\lambda$ denote the
$f_k^N+1^{\text{st}}$ largest eigenvalue of $L_k$.  Then for any $c>1$,
\be 
\label{thmeq}
\Pr\Bigl[\lambda \geq c \Bigl( \lambda_H^{\frac{1}{k+1}} + 
O(\frac{\log(k)+\log(\log(N))}{\log(N)})\Bigr)\Bigr] \leq
c^{(-k+1)\log_{1/\lambda_H}(N)},
\ee 
\end{theorem}
where $\Pr[...]$ denotes probability and $\lambda_H$ depends on $D$
as given in Eq.~(\ref{lHdef})..

Note that since $\lambda_H^{1/(k+1)}$ converges to unity as $k$ becomes
large, the result (\ref{thmeq}) is only meaningful for $k=\cO(\log(N)/\log(\log(N)))$.
Constants depending on $D$ are also hidden inside of the
$O()$ notation.  The result is likely far from optimal, since numerical
studies indicate that for fixed $k$ and large $N$, the largest non-trivial
eigenvalue
$\lambda$ approaches $\lambda_H$.
This result for the case $k=1$ was only recently
proven\cite{friedman}.
Our proof, which gives a weaker bound on the expectation value
of $\lambda$ 
roughly follows the presentation of the trace method
in \cite{expanderbook,bs}, with some modifications.

{\em Proof of \thm{random-cl-TPE}:}
We will apply the trace method separately
in each of the subspaces $S_a$.
It suffices to consider only one such subspace $S_a$, the subspace
$S_{f_k^N}$ in which
all of the $n_1,n_2,\ldots,n_k$ differ from each other, since every
eigenvalue
of $L_k$ is an eigenvalue of $L_k$ restricted to $S_{f_k^N}$.
For example, consider the case $k=2$.  We have
two different subspaces, one with $n_1=n_2$ and one with $n_1\neq n_2$.
The eigenvectors of the first subspace, of the form
$\sum_i p(i) |i\rangle |i\rangle$, correspond to eigenvectors of $L_1$ of the
form $\sum_i p(i) |i\rangle$.
Given such an eigenvector, we can construct an eigenvector in the
second subspace equal to $\sum_i \sum_{j\neq i} p(i) |i\rangle|j\rangle$
with the same eigenvalue, as claimed.

Let $\bbE[...]$ denote an average over different choices of permutation
matrices.  Then for any even $m$,
\be
\label{ctm}
\bbE[|\lambda|]\leq (\bbE[{\rm tr}(L_k^m R)]-1)^{1/m},
\ee
where $R$ is the projector onto the given subspace.
The expectation value $\bbE[{\rm tr}(L_k^m R)]$ equals
\be
\Bigl(\frac{1}{D}\Bigr)^m \sum_{s_1=1}^D \sum_{s_2=1}^D ...
\sum_{s_m=1}^D 
\bbE[{\rm tr}(P(s_1) P(s_2) ... P(s_m) R)].
\ee
If for some $i$ we have $s_i=s_{i+1}+D/2$, then $P(s_i)P(s_{i+1})=\openone$,
and we can remove that pair of permutation matrices from the trace above.
Similarly, if $s_m=s_1+D/2$, then we can remove the first and last permutation
matrices from the trace, exploiting the cyclic invariance of the trace
and the vanishing commutator $[P(s),R]=0$.
We can consider these operations as acting on a word
$s_1,s_2,...,s_m$  on an alphabet
$\{1,...,D\}$.  We define a reduced word
by removing pairs of letters of the form $s,s+D/2$.  Similarly,
if the word ends with a letter $s$ and begins with a letter $s+D/2$, we
remove this pair also.  We repeat these removals until no further removals
are possible.  The result is a reduced word of length $m^0\leq m$;
the resulting sequence we write $s'_1,s'_2,...,s'_{m^0}$.
There are at most 
\be
\label{atmost}
(D-1)^{m/2} 2^m=D^m \lambda_H^m
\ee
choices of $s_1,...,s_m$ which
give $m^0=0$; the number of these choices is equal to
$D^m$ times
the return probability of a random walk of length $m$ on
a Cayley tree of degree $D$.
For these choices, we have
$\bbE[{\rm tr}(P(s_1) P(s_2) ... P(s_m) R)]={\rm tr}(R)\leq N^k$.

We now consider the other choices of $s_1,...,s_m$, where $m^0>0$.
In general, 
\be
\bbE[{\rm tr}(P(s'_1) P(s'_2) ... P(s'_{m^0}) R)] \leq
N^k
\bbE[{\rm tr}(P(s'_1) P(s'_2) ... P(s'_{m^0}) R_{1,2,...,k})],
\ee
where
$R_{1,2,...,k}$ projects onto the state with $n_1=1,n_2=2, ...,n_k=k$.
To compute this expectation value, we define $v_0^a=a$, for $1\leq a \leq k$.
Then, define $v_i^a$, for $i\geq 1$ and $1\leq a \leq k$, to be
$\pi_{s'_i}(v_{i-1}^a)$.  
Then, the probability that $v_{m^0}^a=a$ for all $a$
is equal to the desired result.
We compute this probability as follows.
Consider this as happening sequentially, where
first we define $v_1^a$ for all $a$, then we define $v_2^a$, and so on.
We say that a choice of $v_i^a$ is ``free'' if at no previous step $j<i$
did we compute $\pi_{s'_j}(v_{j-1}^b)$ with $s'_j=s'_i$ and
$v_{j-1}^b=v_{i-1}^a$.  If a choice of $v_i^a$ is free, and if
$t$ values of $\pi_{s'_i}$ have been previously revealed, than we can
simply pick $v_i^a$ at random from the $N-t$ possibilities, thus revealing
some of the information about the permutation $\pi_{s_i^a}$, and increasing
$t$ by one for that permutation.
If a choice is not free, then it is ``forced'', in which case we have
no choice about the value of 
$\pi_{s'_i}(v_{i-1}^a)$.

We say that a coincidence occurs at step $i$ for walker $a$ if
this is a free step and the randomly selected vertex coincides with
a previously selected vertex (previously selected by {\it any} of the walkers).
Note that for 
$v_{m^0}^a$ to equal $a$ for all $a$, we must have at least $k$ coincidences.
There are two cases: either there are at least $k+1$ coincidences, or else
there are exactly $k$ coincidences.

The probability of there
being at least $k+1$ coincidences can be computed as follows.
Let $i_1,i_2,...,i_{k+1}$ be the steps of the first $k+1$ coincidences
and $a_1,a_2,...,a_{k+1}$ be the corresponding walkers.  The probability
of having these coincidences for given $i_1,...$ and $a_1,...$ is bounded
by
$(mk/(N-mk))^{k+1}$.  Summing over all possible steps and walkers, we find
that the probability of having at least $k+1$ coincidences is bounded by
\be
\label{plots}
m^{k+1} k^{k+1} (mk/(N-mk))^{k+1}.
\ee

If there are exactly $k$ coincidences, then each walker has exactly
one coincidence given that $v_{m_0}^a=a$ for all $a$.
There are two possibilities:
either all of the coincidences occur on the last step, or at least
one coincidence does not occur on the last step.  
The probability of the
first case is at most $(1/(N-mk))^k$.  If at least one coincidence does
not occur on the last step, then let walker $b$ be the first walker
to have a coincidence, occurring on step $j$.  
Note that
each of the vertices $1,...,a$ must be the randomly selected vertex
on exactly one coincidence, again given that $v_{m_0}^a=a$ for all $a$.
Because there are no further coincidences for walker $b$, we have
$s'_i=s'_{i+j}$ for all $i$.  The fraction of reduced words of length $m_0$
that obey this constraint for given $j\leq m_0/2$ is at most $(D-1)^{-m_0/2}$.
The fraction of words that have a reduced word of length $m_0$ is
at most $(D-1)^{m_0/2}\lambda_H^m$.  Therefore, the fraction of words
that have a reduced word obeying this constraint, after summing over $j$,
is at most $m \lambda_H^m$.
The probability of having these coincidences is bounded by
$(m/(N-mk))^k$, where the factor of $m$ arises from the choice of
step on which the coincidence occurs (this is in fact a large overestimate).
The product of these probabilities is $m\lambda_H^m 
(m/(N-mk))^k$.
The total of these two possibilities is
\be
\label{pco}
(1/(N-mk))^k+(m/(N-mk))^k m\lambda_H^m).
\ee

Adding 
the sum of the expectation
value over words with $m^0=0$ (which is bounded by $N^k \lambda_H^m$ by
Eq.~(\ref{atmost}) to
$N^k$ times the sum of
(\ref{plots},\ref{pco}), we find that
\be
\bbE[{\rm tr}(P(s'_1) P(s'_2) ... P(s'_{m^0}) R)] \leq
N^k \lambda_H^m+
N^k m^{k+1} k^{k+1} (mk/(N-mk))^{k+1}+
(N/(N-mk))^k+(Nm/(N-mk))^k m\lambda_H^m.
\ee
and therefore
\begin{eqnarray}
&& \bbE[{\rm tr}(P(s'_1) P(s'_2) ... P(s'_{m^0}) R)]-1 
\\ \nonumber
&\leq&
N^k \lambda_H^m+
N^k m^{k+1} k^{k+1} (mk/(N-mk))^{k+1}+
[(N/(N-mk))^k-1]+(Nm/(N-mk))^k m\lambda_H^m \\ \nonumber
&=&
N^k \lambda_H^m+
N^k m^{k+1} k^{k+1} (mk/(N-mk))^{k+1}+
\cO(mk^2/N)+(Nm/(N-mk))^k m\lambda_H^m.
\end{eqnarray}
We pick
\be
m=(k+1) \log_{1/\lambda_H}(N)
\ee
to minimize this expectation value, finding
\be
(\bbE[{\rm tr}(P(s'_1) P(s'_2) ... P(s'_{m^0}) R)]-1)^{1/m} \leq
\lambda_H^{1/(k+1)} (\cO(mk))^{(k+1)/m}.
\ee
Applying Markov's inequality then yields the proof of the Theorem.\qed

\subsection{Quantum expanders from classical tensor product expanders}
One application of $k=2$ classical tensor product expanders is to
constructing quantum expanders.  We give two constructions.

The first approach was introduced, but not formally analyzed, in
\cite{entqgs}. 
Let $P(s)$ be a set of random permutation matrices defining a $k=2$
tensor product expander, as in the random construction of a $k=2$
tensor product expander above.
Then, define $\sigma(s)$, for $s=1...D$, to
be a diagonal matrix.  For $s=1,...,D/2$ we
choose $\sigma(s)$ to have diagonal entries $\pm 1$ chosen independently
at random and we choose
$\sigma(s+D/2)=P(s)\sigma(s) P(s)^{\dagger}$.
Then, in \cite{entqgs} it was shown numerically that the
$A$ matrices,
\be
\label{aA}
A(s)=\frac{1}{\sqrt{D}} P(s) \sigma(s),
\ee
define a quantum expander with high probability.
Note that the choice of $\sigma(s+D/2)$ is such that
$A(s+D/2)=A(s)^{\dagger}=(1/\sqrt{D}) \sigma(s)P(s)^{\dagger}$
 so that this is a Hermitian expander
because $P(s)=P(s+D/2)^\dagger$.
Numerically,
$\lambda$ was observed to approach
$\lambda_H$ for large $N$.  
We now prove that we do indeed get a quantum expander with
high probability, but
with a weaker bound on $\lambda_H$.
\begin{theorem}
Choose $\pi_1,\ldots,\pi_{D/2}\in\cS_N$ at random and then take
$\pi_{s+D/2}=\pi_s^{-1}$.  Let $P(s)=P_{\pi_s}$.
Choose $\sigma(s)$ as described above.  Let $\lambda$ denote
the second largest eigenvalue of the map with
Kraus operators given by the matrices $A(s)$ in
Eq.~(\ref{aA}).
Then,
for any $c>1$,
\be
\Pr\Bigl[\lambda \geq c \Bigl( \lambda_H^{\frac{1}{3}} + 
O(\frac{\log(\log(N))}{\log(N)})\Bigr)\Bigr] \leq
c^{-3\log_{1/\lambda_H}(N)}.
\ee
\end{theorem}

The Hermitian,
completely positive map ${\cal E}$ defined by the $A$ matrices 
in (\ref{aA}) sends
a diagonal matrix to a diagonal matrix and an off-diagonal matrix to
an off-diagonal matrix.  So, we consider the spectrum of ${\cal E}$ in the
diagonal and off-diagonal sectors separately.  In the diagonal sector,
the spectrum of ${\cal E}$ is the same as that of the $k=1$ expander
defined by the given permutation matrices, and hence has a gap between the
largest eigenvalue, equal to unity, and the next largest eigenvalue.

The off-diagonal sector requires a little more work.  We
again use the trace method.  Let $\lambda$ be the largest eigenvalue
in absolute value in the off-diagonal sector.  Let $M(i,j)$ be an $N$-by-$N$
dimensional matrix with a one in the $i^{\text{th}}$ row and $j^{\text{th}}$ column, and zeroes
everywhere else, so that these form a basis for the space of $N$-by-$N$
matrices.   The $M(i,j)$ with $i\neq j$ form a basis for the space of
off-diagonal matrices.
Define $(M,N)$ to be an inner product on the space of $N^{k}$-by-$N^k$
dimensional matrices by $(M,N)={\rm tr}(M^\dagger N)$.
Then for any even $m$,
\be
\label{qefromct}
\bbE[|\lambda|] \leq
\Bigl( \bbE[\sum_{i\neq j} \Bigl(M(i,j),{\cal E}^m(M(i,j))\Bigr )]\Bigr)^{1/m}.
\ee
Note that compared to Eq.~(\ref{ctm}), a factor of unity is not
subtracted from the expectation value on the right-hand side of
Eq.~(\ref{qefromct}).

The evaluation of the right-hand side of Eq.~(\ref{qefromct}) proceeds
analogously to that of Eq.~(\ref{ctm}).  The computation in the case
$m^0=0$ is identical.  In the case $m^0>0$, we again define coincidences
and paths.  The only difference is that now rather than just computing
the probability that $v_{m^0}^a=a$ for all $a=1,2$, the paths come in
with signs which may be plus or minus one.  This can only reduce the
contribution of the terms with $m^0>0$.  We bound the case with $k+1$
coincidences as before.  We also bound the case with $k$ coincidences
not all occurring on the last step
as before.  The only difference is the case in which all coincidences happen
on the last step $i=m^0$.  The probability of this happening is $(1/N)^2$.  The
sign, however, is completely random; it is equally likely to be plus or
minus one.  Thus, the paths with exactly $k$ coincidences, all occurring
on step $i=m^0$, contribute zero to the expectation value (\ref{qefromct}).
Thus, 
\be
\bbE[|\lambda|^m] \leq
(N^k+m) \lambda_H^m+
N^k m^{k+1} k^{k+1} (mk/(N-mk))^{k+1}.
\ee
Picking $m$ as before, we find that $\bbE[|\lambda|]\leq 
\lambda_H^{1/3}
(1+\cO(\log(\log(N))/\log(N))$.
Applying Markov's inequality yields the theorem.
\qed

We now describe our second construction of a quantum expander from a
classical tensor product expander.
\begin{theorem}\label{thm:2c-1q}
Suppose $\{P(1),\ldots,P(D)\}$ form a
 $(N,D,1-\eps,2)$ classical tensor product expander (i.e. $k=2$).
Assume that $N\geq 2$.
 Let 
$$Z = \sum_{j=1}^{N} \proj{j} e^{\frac{2\pi i j}{N}}$$ and
 $p=1/(1+\eps)$.   Define a quantum operation
 $\cE(M)$ with $D+1$ Kraus operators
 $\sqrt{\frac{p}{D}}P(1),\ldots,\sqrt{\frac{p}{D}}P(D), \sqrt{1-p}Z$.
 Then $\cE$ is a $(N,D+1,1-\frac{\eps}{48})$ quantum expander. 
\end{theorem}

Thus, any constant-gap classical 2-TPE can be used to construct a
constant-gap quantum expander.  No attempt has been made to optimize the constant 48, which we believe can be made arbitrarily close to one when $N$ is large and $\eps$ is close to 1.

Note that $\sqrt{\frac{p}{D}}P(1),\ldots,\sqrt{\frac{p}{D}}P(D), \sqrt{1-p}Z$ is not in general Hermitian, but if $\{P(1),\ldots,P(D)\}$ is Hermitian then 
$\{\sqrt{\frac{p}{D}}P(1),\ldots,\sqrt{\frac{p}{D}}P(D), \sqrt{\frac{1-p}{2}}Z,
\sqrt{\frac{1-p}{2}}Z^\dag\}$ is a Hermitian $(N,D+2,1-\eps/48)$ expander; this is proved by using the triangle inequality to relate its gap to the gap of the expander in \thm{2c-1q}.

{\em Proof of \thm{2c-1q}:}
The idea is that the classical TPE randomizes the diagonal elements of
the density matrix simply because it is an expander, and it randomizes
the
off-diagonal elements because it is a $k=2$ TPE. Next the phase
operation $Z$ adds a phase to the off-diagonal elements so that they
are no longer fixed by the classical TPE.  Thus the only fixed state
will be the identity matrix.

More formally, let $\ket{\varphi_1}=\frac{1}{\sqrt{N}}\sum_{i=1}^N
\ket{i}\ket{i}$ and
$\ket{\varphi_2}=\frac{1}{\sqrt{N(N-1)}}\sum_{i\neq j}
\ket{i}\ket{j}$.  These two states form an  orthonormal basis for the
invariant subspace of $\frac{1}{D}\sum_{s=1}^D P(s)\ot P(s)$. Thus the
fact that $P(1),\ldots,P(D)$ form a 2-TPE implies the bound

$$ \l\|\frac{1}{D}\sum_{s=1}^D P(s)\ot P(s) - \varphi_1 -\varphi_2\r\|
\leq\lambda.$$
Next, a short calculation shows that
$\bra{\varphi_2}Z\ket{\varphi_2}=-1/(N-1)$.  Now apply the 
following Lemma to the subspace orthogonal to $\ket{\varphi_1}$.

\begin{lemma}\label{lem:mix-projs}
Let $\Pi$ be a projector and let $X$ and $Y$ be operators such that
$\|X\|\leq 1$, $\|Y\|\leq 1$, $\Pi X = X \Pi = \Pi$,
$\|(I-\Pi)X(I-\Pi)\| \leq 1 - \eps_X$ and 
$\|\Pi Y \Pi\| \leq 1 - \eps_Y$.  Assume $0<\eps_X, \eps_Y<1$.
Then for any $0<p<1$, $\|p X + (1-p)Y\| < 1$.  Specifically,
\be \| p X + (1-p)Y \| \leq 
 1-\frac{\eps_Y}{12}\min(p\eps_X,1-p).
\label{eq:intermediate-norm}\ee
Setting $p = 1 / (1 + \eps_X)$, we obtain
\be \| p X + (1-p)Y \| \leq 1 - \frac{\eps_X\eps_Y}{12(1+\eps_X)}
\leq 1 - \frac{\eps_X\eps_Y}{24}.
\label{eq:mixed-norm}\ee
\end{lemma}
The Lemma is proved in Appendix~\ref{app:lem-mix}.
We apply the Lemma by taking $X = \frac{1}{D}\sum_{s=1}^D P(s)\ot P(s)
- \varphi_1$,
$Y=Z\ot Z^* - \varphi_1$ and $\Pi=\varphi_2$.
Then plugging $\eps_X=\eps$ and $\eps_Y=1-1/(N-1)\geq 1/2$
into \eq{mixed-norm} completes the proof of
\thm{2c-1q}.\qed

\section{Quantum Tensor Product Expanders}
\label{sec:quantum-TPE}

In this section we define quantum tensor product expanders
and show that random
unitaries provide a way of constructing tensor product expanders.  
We begin with some preliminaries and definitions, present
applications to the Solovay-Kitaev problem of approximating unitaries by a
string of elementary operations, and finally prove
that random unitaries give tensor product expanders.
The proof of this last statement
begins in subsection \ref{sec:TPE-proof};
it closely follows \cite{sd} and should be read in conjunction
with that paper.

\subsection{Preliminaries, Definitions, and Applications}
Suppose we
have a collection of unitaries $\{U(1),\ldots,U(D)\}\in \cU_N$.  Define
a quantum operation $\cE_k$ that applies  $U_s^{\ot k}$ for
$s\in\{1,\ldots,D\}$ chosen uniformly at random.  In other words
\be 
\label{calEdef}
\cE_k(M) = \frac{1}{D} \sum_{s=1}^D U(s)^{\ot k} M
(U(s)^\dag)^{\ot k},
\ee
where $M$ is an $N^k\times N^k$ matrix.  Since an $N^k\times N^k$
matrix can also be viewed as an $N^{2k}$-dimensional vector, we
can also interpret $\cE_k$ as a linear operator on an
$N^{2k}$-dimensional vector space.  Define this operator to be
\be \hat{\cE}_k := \frac{1}{D} \sum_{s=1}^D U(s)^{\ot k}
\ot (U(s)^*)^{\ot k}.\ee
Note that $\cE_k$ and $\hat{\cE}_k$ are isospectral.

In previous work\cite{sd,qexp,qexp2} $\cE_1$ was said to be a
$(N,D,\lambda)$ quantum expander if the second-largest eigenvalue
of $\hat{\cE}_2$ was $\leq \lambda$.  In fact, the definition of
quantum expanders included even quantum operations that were not
mixtures of unitaries, as long as they could be expressed using $\leq
D$ Kraus operators.  Here we will change notation from
\cite{sd,qexp,qexp2} slightly. We say that a set of unitaries
$\{U(1),\ldots,U(D)\}$ is a $(N,D,\lambda,k)$ tensor product expander
if the operator ${\cE}_k$ has $F_k^N$ (defined below) eigenvalues
equal to one, and all of its other eigenvalues have absolute value
$\leq \lambda$.  This differs from the notation of
\cite{sd,qexp,qexp2} in that the set of unitaries, rather than the
quantum operation, constitutes the quantum expander\footnote{One can slightly generalize this by defining a set of
unitaries and a set of associated probabilities to be a tensor product
expander; however in this paper we consider applying each unitary with equal probability summing to
unity.}.
When $N$ and $D$
are understood, we sometimes simply say that $\{U(1),\ldots,U(D)\}$
are a $k$-tensor product expander with gap $1-\lambda$.

We define $F_k^N$ to be the rank of the projector
$$\hat{\cT}_k := \int_{U\in \cU_N} U^{\ot k} \ot (U^*)^{\ot k} \d U$$ or
equivalently of the operation $\cT_k$, which is defined by 
\be \cT_k(M) =
\int_{U\in \cU_N} U^{\ot k} M (U^\dag)^{\ot k}.
\label{eq:twirl-def}\ee  (Throughout the
paper the integration measure ${\rm d}V$ will be the Haar measure.)
This map is the ``twirling'' operation\cite{twirl}.  Since $\cT_k$ is
a Hermitian map and $\cT_k(\cT_k(M))=\cT_k(M)$, the map $\cT_k(M)$ has
all eigenvalues equal to zero or unity.

 For $\pi\in \cS_k$, we define the $N^k\times N^k$ matrix $\bP_N(\pi)$ is
defined to be
$$\bP_N(\pi) = \sum_{i_1=1}^N \cdots \sum_{i_k=1}^N \ket{i_1,\ldots,i_N}
\bra{i_{\pi(1)},\ldots,i_{\pi(N)}}.$$ Since $\bP_N(\pi)$ commutes with any
matrix of the form $U^{\ot k}$, it follows that
$\cT_k(\bP_N(\pi))=\cE_k(\bP_N(\pi)) = \bP_N(\pi)$ for any $\pi$.  We claim that
the $\bP_N(\pi)$ (and their linear combinations) constitute all of the
unit eigenvalues of $\cE_k$.  This fact follows from Schur-Weyl
duality, and specifically Thm 3.3.8 of \cite{GW} which states
that $\cT_k(M) = M$ if and only if $M$ is a linear combination of
$\bP_N(\pi)$ operators.  Thus $F_k^N = \dim\Span\{\bP_N(\pi) : \pi\in
\cS_k\}$.

An important special case is when $N\geq k$.  In this case,
 the set $\{\bP_N(\pi)\ket{1,2,\ldots,k}:\pi\in\cS_k\}$ is linearly
 independent, which implies that $\{\bP_N(\pi):\pi\in\cS_k\}$ is linearly
 independent and thus that
$F_k^N=k!$.

In the quantum case, tensor product expanders give us a way to
approximate the twirling operator $\cT_k$ of \cite{twirl}.  This is
because 
\be \|\cE_k^m - \cT_k\|_\infty \leq \lambda^m,\ee
so whenever $\lambda<1$, $\cE_k^\infty = \cT_k$. Let us consider
various other possibilities for implementing twirling as a sum of
different unitary transformations: one approach to exactly
implementing the twirling operation is to use
$t$-designs\cite{tdesigns}, but the number of unitaries that must be
implemented in this case grows with $N$.  Another approach was
discussed in \cite{gt}, which avoids having the number of unitaries
grow in $N$, but requires the ability to implement a number of
unitaries growing linearly in the logarithm of the error of the
approximation.  In contrast, tensor product expander require only the
ability to implement a constant number of unitaries to get arbitrarily
good approximations.  This is a definite advantage; however, in
practice, our construction of tensor product expanders here, which
relies on the ability to construct random unitary operations, probably
cannot be efficiently implemented using gates; instead, we would like
to efficiently implement a deterministically constructed tensor
product expander.  This raises the interesting question of whether the
constructions of \cite{qexp2} can lead to tensor product expanders
also.

{\em The limit of large $k$:}  The situation when $k$ is large has
some similarities to the classical case.  It still holds that any
$(N,D,\lambda,k)$ quantum tensor product expander is also a $(N,D,\lambda,k')$ quantum
tensor product expander for all $k'\leq k$.  In particular, if a set
of unitaries forms a $(N,D,\lambda,\infty)$ quantum tensor product
expander than it is also a $(N,D,\lambda,k)$ quantum tensor product
expander for any finite $k$.  This is equivalent to generating a
Cayley graph expander on $\cU_N$.  One difference between the quantum
and classical cases is that there is no upper bound to the size of
irreps of $\cU_N$, like there is for $\cS_N$.

Note that constant degree Cayley graph expanders are known for
$\cU_2$; indeed, choosing the matrices at random will yield an
expander with probability one\cite{SU2-expand}.  However, no proof of this fact is
known for $N>2$.
\subsection{Solovay-Kitaev gate approximation}

One application of tensor product expanders is to the problem of
approximating an arbitrary $V\in\cU_N$ with a string of gates from a
fixed universal set $\{U(1),\ldots,U(D)\}$.  The fact that
$\{U(1),\ldots,U(D)\}$ is universal means that $\<U(1),\ldots,U(D)\>$ is
dense in $\cU_N$ (optionally neglecting an overall phase).  This means
that for any $V\in\cU_N$ and any $\eps>0$, there exists a string
$s_1,\ldots,s_m$ such that $U(s_1)U(s_2)\cdots U(s_m)$ is within
a distance $\eps$ of $V$.  Often we also
want to know (a) how quickly $m$ grows with $1/\eps$ and (b) how
long it takes to find $s_1,\ldots, s_m$.  When $\{U(1),\ldots,U(D)\}$
contain their own inverses, the Solovay-Kitaev theorem\cite{sk} gives
a $\poly\log(1/\eps)$ time (for fixed $N$) algorithm to find an
$\eps$-approximation with $m = O(\log^{3+o(1)}(1/\eps))$.
Very little is known in the case without access to inverses,
except that $U(s)^\dag$ can be simulated to error $\eps$ using
$O(1/\eps^{N^2})$ applications of $U(s)$, meaning that the
Solovay-Kitaev construction can be used with this amount of overhead.

Turning to lower bounds, observe a ball of radius $\eps$ in $\cU_N$ has volume
$\Theta(\eps^{N^2})$.  This implies that to approximate all strings to
within error $\eps$ requires $\Omega((1/\eps)^{N^2})$ different unitaries,
or equivalently a $\Omega(N^2\log 1/\eps)$ string length.
A long-standing open question is whether the Solovay-Kitaev
approximation can in general be improved to use the optimal $O(\log
1/\eps)$ number of gates.  Such optimally short approximations are
known to exist whenever a particular random walk on $\cU_N$ has a
gap\cite{HRC}: specifically, the walk consisting of multiplying by
$U(s)$ for $s$ randomly chosen from $1,\ldots,D$.  For $\cU_2$, it
was recently proven that generic $U(1),\ldots,U(D)$ are
gapped\cite{BG} and thus yield short approximating strings. However,
the situation for $\cU_N$ for $N>2$ remains open.

 In this section we will prove that when  $k$ is sufficiently large, 
unitaries forming $k$-tensor product expanders yield optimal
$O(N^2\log 1/\eps)$-length $\eps$-approximations for any gate in
$\cU_N$.
\begin{theorem}. \label{thm:SK-tpe}
 Suppose
$\{U(1),\ldots,U(D)\}$ form a $k$-tensor product expander with gap
$1-\lambda$ for $k \gg
\frac{N^3\log^2(1/\eps)}{\eps}$.  Then for any
$V\in\cU_N$ there exists a string $s_1,\ldots,s_m\in\{1,\ldots,D\}$ with
$m=O(N^2\log_{1/\lambda} (1/\eps))$  and $d(V, U(s_1)U(s_2)\cdots
U(s_m)) \leq \eps$.
\end{theorem}
Here we define the distance between two unitaries $d(U,V)$ by
$$d(U,V) = \min_{\phi\in[0,2\pi]} \| U - e^{i\phi} V\|_2 = 2N - 2|\tr U^\dag
V|,$$ so that it ignores overall phase.  

The main result from \cite{HRC} can be thought of a
weaker version of \thm{SK-tpe}: it requires $k=\infty$ to achieve the same
conclusion.  Unfortunately, \thm{main} only shows that generic
sets of unitaries are $k$-tensor product expanders for $k\sim
N^{1/6}/\log(N)$.  Thus, at present the existence of expanders
satisfying the assumptions of \thm{SK-tpe} is a nontrivial conjecture.
It is possible that there exists some
strengthening of the results of \thm{main} which will allow us to
show that generic unitaries fulfill the assumptions of \thm{SK-tpe}.

{\em Proof of \thm{SK-tpe}:}
Let $\ket{\Phi}=\frac{1}{\sqrt{N}}\sum_{i=1}^N \ket{i}\ket{i}$ be the
maximally entangled state 
on $\bbC^N \ot \bbC^N$.  Define $\rho(U) = \l[(U\ot I)\Phi(U^\dag \ot
I)\r]^{\ot k}$.
Observe that 
\be \tr \rho(U)\rho(V) = |\tr U^\dag V|^{2k} / N^{2k}
= \l(1 - \frac{d(U,V)}{2N}\r)^{2k}
\label{eq:rho-dist}\ee

Let $B_{\epsilon/3}$ be the ball of radius $\epsilon/3$ around the
identity: $B_{\epsilon/3}=\{U| d(U,I) \leq \epsilon/3\}$.  Let
$\Vol(\epsilon/3)$ denote the volume of $B_{\epsilon/3}={\cal
O}((\epsilon/3)^{N^2})$.  Define
\be
\rho_{\epsilon}(U)= \frac{1}{\Vol(\eps/3)}
\int_{V\in B_{\epsilon/3}} \rho(VU) {\rm d}V,
\label{eq:rho-eps-def}
\ee

Similarly we define 
\be\rho_H = \int_{V\in \cU_N} \rho(V) {\rm d}V.
\label{eq:rho-H-def}\ee
These states are normalized so that $\tr
\rho_\eps(U) = \tr \rho_H = 1$.  Since $\rho(V)\geq 0$ for all $V$, we have the operator inequality $\rho_\eps(U)
\leq \rho_H / \Vol({\eps/3})$
for any $U$.
Also observe that
$\rho_H = (\cT_k\ot \text{id}_N^{\ot k})(\rho(U))$ for any $U$, where $\text{id}_N$ denotes the identity operation on $N\times N$ density matrices.

We will find it convenient to think of density matrices as vectors with the Hilbert-Schmidt inner product $\<A,B\> = \tr A^\dag B$.  In this picture $\cT_k$ is a projector, and so
$$\tr \rho_\eps(U)\rho_H = 
\tr \rho_\eps(U)(\cT_k\ot \text{id}_N^{\ot k})(\rho_\eps(U)) = \tr \rho_H^2.$$
To bound $\tr \rho_H^2$, observe that the support of $\rho_H$ lies within
$\Span\{\ket{\psi}^{\ot k} : \ket{\psi}\in\bbC^{N^2}\}$, which (according to \cite{sym, GW}) has
dimension $\binom{N^2+k-1}{N^2}  = k(k+1)\cdots (k+N^2-1)/N^2! \leq k^{N^2}$.  Thus $\tr \rho_H^2 \geq k^{-N^2}$.

Now we use the fact that $\|\cE_k^m - \cT_k\|_\infty \leq \lambda^m$
together with Cauchy-Schwartz to bound
\be\tr \rho_\eps(I) \cE^m(\rho_\eps(U)) \geq \tr \rho_\eps(I)\rho_H
- \lambda^m \tr\rho_\eps(I)^2 
\geq \tr\rho_H^2\l(1- \frac{\lambda^m}{\Vol({\eps/3})^2}\r)
\geq \frac{1}{2}\tr\rho_H^2 \geq \frac{1}{2k^{N^2}},
\label{eq:rho-well-mixed}\ee
where in the second-to-last step we have assumed $m \geq
\log(2/\Vol({\eps/3})^2) / \log(1/\lambda) = \cO(N^2
\log_{1/\lambda}(1/\eps))$.

On the other hand, if there is no string $s_1,...,s_m$ such that
$d(U(s_1) U(s_2) ... U(s_m),  U) \leq \epsilon$,
then
\be \tr \rho_\eps(I) \cE^m(\rho_\eps(U))  \leq
\l(1 - \frac{\eps}{6N}\r)^{2k} \leq e^{-\frac{k\eps}{3N}}.
\label{eq:all-strings-far}\ee

If $k/\log k\gg N^3/\eps$ then \eq{rho-well-mixed} and
\eq{all-strings-far} cannot simultaneously hold.  Therefore there must
exist at least one string $s_1,\ldots,s_m$ for which 
$d(U(s_1) U(s_2) ... U(s_m),  U) \leq \epsilon$.
\qed

\subsection{Trace Method and Schwinger-Dyson Equations}
\label{sec:TPE-proof}
The next three sections are devoted to the expansion properties
of randomly chosen unitaries.
Recall that we would like to construct a quantum tensor product
expander by randomly choosing $U(1),\ldots,U(D)\in\cU_N$.
There are two cases.  In the non-Hermitian case, the unitary matrices
$U(s)$ are chosen independently with the Haar measure.  In the Hermitian
case, $D$ is even and the unitary matrices $U(s)$ for $s=1,\ldots,D/2$ are
chosen independently with the Haar measure and $U(s+D/2)=U(s)^{\dagger}$,
so that ${\cal E}_k$ is a Hermitian operator.  We focus on the Hermitian
case, and the techniques can be readily extended to cover the
non-Hermitian case.  Our main result is that for random $U(s)$, with high probability
we do indeed get a tensor product expander:
\begin{theorem}. \label{thm:main}
Let $\{U(1),\ldots,U(D/2)\}$ be chosen randomly with the Haar measure
from the unitary group $\cU_N$, and let $U(s+D/2)=U(s)^{\dagger}$. 
Let $k\leq\cO(N^{1/6}/\log(N))$ and
let $\lambda$ denote the $F^N_k+1^{\text{st}}$ eigenvalue of 
${\cal E}_k$ as defined in (\ref{calEdef}).
Then, for any $c>1$,
\be
\Pr\l[\lambda\geq
c(1+\cO(k\log(N) N^{-1/6})\lambda_{H}\r] \leq
c^{-(1/4k) N^{1/6}},
\ee
where $\lambda_H$ depends on $D$ and is given in Eq.~(\ref{lHdef}).
\end{theorem}

We use a trace method to bound the eigenvalues of
${\cal E}_k(M)$.
We have
\be
\label{tr1}
\sum_{i_1,i_2,...,i_k}\sum_{j_1,j_2,...,j_k} \Bigl(
M(i_1,j_1)\otimes
M(i_2,j_2)\otimes ... 
\otimes
M(i_k,j_k),
{\cal E}_k^m(M(i_1,j_1)\otimes
M(i_2,j_2)\otimes ... 
\otimes
M(i_k,j_k))\Bigr)=\sum_{a=1}^{N^{2k}} |\lambda_a|^m \geq
k!+|\lambda|^m,
\ee
where we pick $m$ to be an even integer.
We will derive bounds on the expectation value of the trace to bound the
expectation of
$|\lambda|^m$.
Eq.~(\ref{tr1}) can be re-written as
\begin{eqnarray}
\label{e0for1}
k!+|\lambda|^m
\leq
\Bigl(\frac{1}{D}\Bigr)^m \sum_{s_1=1}^D
\sum_{s_2=1}^D...
\sum_{s_{m}=1}^D
{\rm tr}(U(s_m+D/2) ...
U(s_2+D/2) U(s_1+D/2))^k
{\rm tr}(U(s_1) U(s_2) ... U(s_m))^k].
\end{eqnarray}

Let $\bbE[...]$ denote the average over the unitary group.
Averaging Eq.~(\ref{e0for1}) we find
\be
\label{av}
E_{1,k}\equiv \Bigl(\frac{1}{D}\Bigr)^m \sum_{s_1=1}^D
\sum_{s_2=1}^D...
\sum_{s_{m}=1}^D
E_{0,k}(s_1,...,s_m)
\geq
k!+\bbE[|\lambda|^m],
\ee
\begin{eqnarray}
\label{e0defU}
E_{0,k}(s_1,...,s_m) &\equiv &
\bbE[
{\rm tr}(U^{\dagger}(s_m) ...
U^{\dagger}(s_2) U^{\dagger}(s_1))^k
{\rm tr}(U(s_1) U(s_2) ... U(s_m))^k]\\ \nonumber &=&
\bbE[
{\rm tr}(U(s_m+D/2) ...
U(s_2+D/2) U(s_1+D/2))^k
{\rm tr}(U(s_1) U(s_2) ... U(s_m))^k].
\end{eqnarray}

As in \cite{sd}, we write the average in Eq.~(\ref{e0defU}) as an average
of the form
\be
\label{prodtraceU}
\bbE[L_1 L_2 ... L_c],
\ee
where
\be
L_1={\rm tr}(U(s_{1,1})U(s_{1,2})...U(s_{1,m_1})), \quad
L_2={\rm tr}(U(s_{2,1})U(s_{2,2})...U(s_{2,m_2})), \quad...
\ee
Here we have an average of $c$ traces, each of which is a product of
some number of unitary matrices.  In particular, Eq.~(\ref{e0defU}) has
$c=2k$, with $L_1=L_2=...=L_k=L_{k+1}^{\dagger}=...=L_{2k}^{\dagger}$.

The Schwinger-Dyson equations for a product of this form are\cite{sd}:
\begin{eqnarray}
\label{SDU}
&&
\bbE[
{\rm tr}(U(s_{1,1})U(s_{1,2})...U(s_{1,m_1}))
L_2 ... L_c]\\ \nonumber
&&=
-
\frac{1}{N}
\sum_{j=2}^{m_1} \delta_{s_{1,1},s_{1,j}}
\bbE[
{\rm tr}(U(s_{1,1})...U(s_{1,j-1}))
{\rm tr}(U(s_{1,j})...U(s_{1,m_1}))
L_2 ... L_c]\\ \nonumber
&&+
\frac{1}{N}
\sum_{j=2}^{m_1} \delta_{s_{1,1},s_{1,j+D/2}}
\bbE[
{\rm tr}(U(s_{1,2})...U(s_{1,j-1}))
{\rm tr}(U(s_{j+1,1})...U(s_{1,m_1}))
L_2 ... L_c]\\ \nonumber
&&-
\frac{1}{N}
\sum_{l=2}^c
\sum_{j=1}^{m_l}
\delta_{s_{1,1},s_{l,j}}
\bbE[
{\rm tr}(U(s_{1,1})...U(s_{1,m_1}) U(s_{l,j}) U(s_{l,j+1}) ...
U(s_{l,j-1}))
L_2 ... L_{l-1} L_{l+1} ... L_c] \\ \nonumber
&&+
\frac{1}{N}
\sum_{l=2}^c
\sum_{j=1}^{m_l}
\delta_{s_{1,1},s_{l,j}+D/2}
\bbE[
{\rm tr}(U(s_{1,2})...U(s_{1,m_1})
U(s_{l,j+1}) U(s_{l,j+2}) ...
U(s_{l,j-1}))
L_2 ... L_{l-1} L_{l+1} ... L_c].
\end{eqnarray}
Note that in the above equation an expression like $U(s_{l,j+1}) U(s_{l,j+2}) ...  U(s_{l,j-1})$ means
$U(s_{l,j+1}) U(s_{l,j+2}) ... U(s_{l,m_l}) U(s_{l,1}) U(s_{l,2}) ...  U(s_{l,j-1})$.

Our general algorithm for reducing traces starts by canceling
all pairs of matrices $U(s) U(s+D/2)$ appearing successively in the
same trace, and replacing ${\rm tr}(\openone)$ by $N$.  We then apply
Eq.~(\ref{SDU}), repeating the cancellation
of successive $U(s) U(s+D/2)$
and replacement of ${\rm tr}(\openone)$ by $N$ on each iteration.
A term terminates at a given level $n$ if there are no matrices left after
$n$ iterations.

Let $m_1^0$ be the length of the trace after canceling
successive $U(s) U(s+D/2)$ before any iterations; on every successive
iteration, the length of the first trace, $m_1$, is bounded by $m_1^0$.
As in \cite{sd}, the number of different choices of $s_1,...,s_m$ which
give rise to a given $m_1^0$ is bounded by
\be
\label{nm1}
(D-1)^{m_1^0/2} (D-1)^{m/2} 2^m.
\ee
This number is equal to $D^m$ times the probability that a random walker
on a Cayley tree arrives at a distance $m_1^0$ from the starting point
after a walk of $m$ steps.
This number is independent of the particular values of
$s_{1,1},...,s_{1,m_1^0}$.  There are $[D/(D-1)] (D-1)^{m_1^0}$ different
possible values of $s_{1,1},...,s_{1,m_1^0}$
and therefore the total number of choices
of $s_1,...,s_m$ which give rise to a given choice of
$s_{1,1},...,s_{1,m_1^0}$ is bounded by
\be
\label{nchoice}
\frac{D-1}{D} \Bigl(\frac{1}{\sqrt{D-1}}\Bigr)^{m_1^0} (D-1)^{m/2} 2^m.
\ee

The number of terms terminating at the $n^{\text{th}}$ level
is bounded by
\be
\label{nterm}
(2km-1)^n.
\ee
To see this, note that at each iteration of the Schwinger-Dyson equation,
the number of terms on the right-hand side is bounded by the number of matrices
on the left-hand side minus one.  Initially, there are $2km$ matrices, and
this number does not increase under Eq.~(\ref{SDU}).

We can estimate the value of a term which terminates at a given level $n>1$
as follows.  First, there is a sign equal to plus or minus $1$.  Next,
there is a factor of $(1/N)^n$.  Finally, there is a factor of $N$ for
each trace of the form ${\rm tr}(\openone)$ that appeared in this process.
Suppose there are $p$ such traces, giving a factor of $N^p$.  How big
can $p$ be?  Initially we have $c=2k$ different traces.  The given term
at level $n$ arose from a specific choice of terms on the right-hand side
of Eq.~(\ref{SDU}) on the first iteration.  This specific choice has $k_1$
different traces in it, with $k_1$ equal to either $k-1$ or $k+1$.
After the second iteration there are $k_2$ traces, then $k_3$, and so on.
The number of traces
$k_2,k_3,...$ can be determined as follows: an application of
Eq.~(\ref{SDU}) may increase the number of traces by one if the term
arises from the first or second line on the right-hand side, or may decrease
the number of traces by one if the term arises from the third or fourth
line on the right-hand side of Eq.~(\ref{SDU}).  Next, some of
the traces may be trivial, being equal to ${\rm tr}(\openone)$.
In the event that the term arose
from the first, second, or third line of Eq.~(\ref{SDU}) it is not possible for
any of the traces to be trivial, under the assumption that any repetitions
of the form $U(s)U(s+D/2)$ have been previously
replaced by $\openone$ in the trace
on the left-hand side of the equation.
However, in the event that the term arose from the fourth line,
then it is possible for one of the traces to be trivial, increasing
$p$ by one.  Thus, for each $b\leq n$,
$k_b-k_{b-1}$ is equal to either $+1,-1,$ or $-2$.
Let $q$ be equal to the number of times the first or second line was used
from Eq.~(\ref{SDU}) and $n-q$ equal the number of times the third or fourth
line was used.  Then, in order for all traces to be trivial in this particular
term resulting from $n$ iterations of Eq.~(\ref{SDU}),
\be
2k+q-(n-q)-p=0.
\ee
Also, since $p$ can only increase when a term from the fourth line is
used,
\be
p \leq n-q.
\ee
Thus,
\be
p \leq \lfloor (2k+n)/3 \rfloor.
\ee
Therefore, the value of  a term terminating
at the $n^{\text{th}}$ level, $n>0$, is bounded in absolute
value by
\be
\label{valBd}
N^{\lfloor (2k+n)/3\rfloor -n}.
\ee
Note that if $m_1^0>0$ then there are no terms terminating at level
$n$ with $n<k$, so for $m_1^0=0$, the trace is equal to $N^{2k}$, while
for $m_1^0>0$, the terms are bound in absolute value by $N^0$ (this bound
is only reached if $k=n$).

Eq.~(\ref{SDU}) generates an infinite series, whose $n^{\text{th}}$
term is the sum of all terms terminating at level $n$.  
As in \cite{sd}, this series is absolutely convergent for
$2km<N$.  In fact, the following stronger claim holds:
Eq.~(\ref{SDU}) generates an absolutely convergent series for $2km-1<N$ which
converges to the expectation value of the trace.
To see this, note that the value $p$ above, the number of traces of
$\openone$, is
always bounded by $2km$.  Thus, the value of a term terminating at
the $n^{\text{th}}$ level is bounded by
\be
\label{pessBd}
N^{2km}N^{-n}.
\ee
Depending on $n$, sometimes (\ref{valBd}) gives a better bound and
sometimes (\ref{pessBd}) gives a better bound, but to estimate convergence
we will use (\ref{pessBd}).  Eq.~(\ref{nterm}) shows that
the number of terms terminating at
level $n$ is bounded by $(2km-1)^n$.
Thus, the
absolute value of the sum of terms terminating at level $n$ is bounded by
$N^{2km} ((2km-1)/N)^n$, and so
for $2km-1<N$, the series is absolutely convergent.  Further,
a term which has not terminated at the $n^{\text{th}}$ level contains at
most $2km$ traces in it, and hence is bounded in
absolute value by $N^{2km} (1/N)^n$.  Therefore,
the sum of all terms which have not terminated at the $n^{\text{th}}$ level
is also bounded by
$N^{2km} ((2km-1)/N)^n)$, and hence for $2km-1<N$ the series converges
to the average of the trace.

\subsection{Example}
We now work out a simple example to give some idea of the use of the
Schwinger-Dyson equations.  This example will also be used later in
the idea of ``complete rung cancellation'' and gives intuition behind
the claim that for $N\geq k$ we have $k!$ eigenvalues equal to unity.
Let the matrix $X$ be chosen from the unitary group with the Haar measure
and evaluate the expectation value for $N\geq k$
\be
\label{exampleX}
\bbE[\Bigl({\rm tr}(X) {\rm tr}(X^{\dagger})\Bigr)^k].
\ee
For $k=1$, a single application of Eq.~(\ref{SDU}) shows that this is equal to
unity.  For $k=2$, we find
\begin{eqnarray}
\label{ex1}
\bbE\l[\Bigl({\rm tr}(X) {\rm tr}(X^{\dagger})\Bigr)^2\r]& = &
2 \bbE\l[\Bigl({\rm tr}(X) {\rm tr}(X^{\dagger})\Bigr)\r]-
(1/N) \bbE\l[{\rm tr}(XX) \Bigl({\rm tr}(X^{\dagger})^2\Bigr)\r] \\ \nonumber
&=& 2+(1/N)^2 \bbE\l[\Bigl({\rm tr}(X) {\rm tr}(X^{\dagger})\Bigr)^2\r]
-2(1/N)^2 \bbE\l[{\rm tr}(X) {\rm tr}(X^{\dagger})\r]\\ \nonumber
&=& 2+(1/N)^2 \bbE\l[\Bigl({\rm tr}(X) {\rm tr}(X^{\dagger})\Bigr)^2\r]
-2(1/N)^2.
\end{eqnarray}
For $N\geq 2$, this shows that
$\bbE[\Bigl({\rm tr}(X) {\rm tr}(X^{\dagger})\Bigr)^2]=2$.

It is interesting to see what happens to the expectation
value in Eq.~(\ref{ex1}) for $N=1,k=2$.  Then,
the last line Eq.~(\ref{ex1}) gives simply 
$\bbE[\Bigl({\rm tr}(X) {\rm tr}(X^{\dagger})\Bigr)^2]=
\bbE[\Bigl({\rm tr}(X) {\rm tr}(X^{\dagger})\Bigr)^2]$, giving no
information about the trace.  For general $N$, the sum of terms terminating at
level $1$ is equal to zero, while the sum of terms terminating at
levels $2,3,4,5,6...$ is equal to $2,-2/N,2/N,-2/N^2,2/N^2,...$ respectively.
Thus, we do not have a convergent series for $N=1,k=2$.

Up to now we have considered the series whose $n^{\text{th}}$ term is the
sum of terms terminating at a given level $n$.  We now consider instead
the expectation value of Eq.~(\ref{exampleX}) as a series in $1/N$.
For $N\geq k$, this series is again absolutely convergent to the
desired expectation value.
It is easy to see that for arbitrary $k$, and for
$N\gg k$, the expectation value (\ref{exampleX})
is equal to $k!+\cO(1/N)$, as there are $k!$ terms which terminate
at level $k$.  We now show that for $N\geq k$, the expectation value
(\ref{exampleX}) is equal to $k!$ exactly.
Note that the expectation value
in Eq.~(\ref{exampleX}) is equal to the trace of the map $\cT_k$
(defined in \eq{twirl-def})

 Thus, the trace of the map $\cT_k(M)$
is equal to the number of unit eigenvalues of $\cT_k(M)$.  For
$N\geq k$ the trace of this map can then be written
as the sum of an infinite series in $1/N$, and using the fact that
the number of unit eigenvalues is equal to an integer for all integer $N$,
we find that all terms in the series in $1/N$, beyond the term of order $N^0$,
must vanish exactly (the calculation above represents an explicit
check of this for $k=2$ and it may be readily verified for any $k$).
Thus, for all $N\geq k$, the expectation value
of Eq.~(\ref{exampleX}) is equal to $k!$.
This gives an alternate proof that $F_k^N=k!$ when $N\geq k$.

\subsection{Counting and Main Result}
In this section we prove a bound on the expectation value of the sum
in Eq.~(\ref{av}), which will give us a bound on the expectation value of
the $m^{\text{th}}$ power of $\lambda$, proving the theorem.  The next
three paragraph are devoted to outlining the basic idea of the
proof, before beginning the technical details.

The basic idea of the
proof is to prove the bound on the sum by proving a bound on the number
of different choices of $s_1,...,s_m$ such that, when the resulting trace
is evaluated using the Schwinger-Dyson equations, there is a term which
terminates at level $n$, for any given $n$.  We give this
bound on the number of choices of $s_1,...,s_m$ in Eq.~(\ref{numnmin}).
We then combine this bound with a bound on the contribution to the trace
of terms which terminate at level $n$.
The idea is that there are a only
small number of choices of $s_1,...,s_m$ which produce terms
which terminates at a small level $n$, and while there are a large
number of choices of $s_1,...,s_m$ which produce terms which terminate
at high levels, such terms are small.

One technical caveat in this work is that for {\it any} choice of $s_1,...,s_m$
there will be certain terms which terminate at a low level $n$.  These
are terms in which we use the Schwinger-Dyson equations to contract
$U(s_i)$ in one trace with $U(s_i)^{\dagger}$ in a different trace.  If
for some $i$, we contract all unitaries $U(s_i)$ in this way,
we have what is called a ``complete rung cancellation" below.  We consider
such terms separately, and they are responsible for producing the
leading order expectation value of the trace in $1/N$: these terms sum to
give a contribution $k!$ to the expectation value of the trace, precisely
corresponding to the expectation we expect from the unit eigenvalues.

Ignoring those terms with complete rung cancellations, we see that a term
in the Schwinger-Dyson equations
must involve contracting $U(s_i)$ with $U(s_j)$ or $U(s_j)^{\dagger}$ for
some $i\neq j$.  Such terms involve constraints: such a term would require
that either $s_i=s_j+D/2$ or $s_i=s_j$.  In order for such a term to terminate
at a low level, there must be many such constraints, and this is why there
are only a few choices of $s_1,...,s_m$ which produce terms which terminate
at low levels.  To show precisely that there are only a few such choices
of $s_1,...,s_m$, we follow a different strategy.  To
explain this strategy,
suppose
you knew a choice of $s_1,...,s_m$ which gave rise to a term which terminated
at some level $n$ and you
were given the task of explaining to someone which choice of $s_1,...,s_m$
you used.  One way to do this would be to simply list the $m$ different
values of $s$.  This would require communicating $\log_2(D^m)$ bits.
We instead
show how to uniquely specify the choices of $s_1,...,s_m$ in a different
way, by specifying most of the choices of $s_1,..,.s_m$ by describing which
cancellations were used.  For small $n$,
this will allow one to communicate the
specific choice of $s_1,..,s_m$ in much shorter way, thus implying that 
that there are only a few choices of $s_1,...,s_m$ which produce the desired
term terminating at level $n$.  We now put this idea into practice.

On a given iteration of the Schwinger-Dyson
equations, we go from a product of $c$ traces to a product of $c+1$, $c-1$,
or $c-2$
traces.  As in \cite{sd}, we keep track of how the matrices move under
this iteration process using a function $f_n((l,i))$ from pairs of integers
to pairs of integers.  We say that the matrix $U(s_{l,i})$ in the
given product of traces, $L_1 L_2 ... L_c$, is in
position $(l,i)$.  Let us consider the case of a term on the first
line, where $c$ increases by one.  Then, for any given $j$ in the
sum on the first line, we say that
the matrix in position $(1,i)$, for $i<j$ on the $n+1^{\text{st}}$ iteration corresponds
to the matrix in position $(1,i)$ on the $n^{\text{th}}$ iteration, and so
$f_n((1,i))=(1,i)$, while the matrix
in position $(2,i)$ on the $n+1^{\text{st}}$ iteration
corresponds to the matrix in position $(1,i+j-1)$ on the $n^{\text{th}}$ iteration, so
$f_n((1,i+j-1))=(2,i)$.
The matrix in position $(l,i)$, for $2<l\leq k+1$ on the $n+1^{\text{st}}$ iteration
corresponds to the matrix $(l-1,i)$ on the $n^{\text{th}}$ iteration, so
$f_n(l-1,i)=(l,i)$.  We follow a
similar procedure for the other lines of Eq.~(\ref{SDU}) and if there are
cancellations, we keep track of how the matrix moves under the cancellations.

We then keep track of which matrix after $n$ iterations corresponds to
a given matrix before any iterations, by defining
$F_n((l,i))=f_n(f_{n-1}(...f_1((l,i))$ for $l=1,2,...,2k$.
Let us say that the matrix
at position $(l,i)$ is ``trivially moved'' under the $n^{\text{th}}$
iteration of the Schwinger-Dyson equations if 
if it is not in either
position $(1,1)$ or position
$(1,j)$ using a term on the first or second line, or in either
position $(1,1)$ or position $(l,j)$
using a term from the third or fourth line.
If a matrix is not trivially moved, and the matrix is not in position
$(1,1)$, then the Schwinger-Dyson equations imply a relation between
$s_{l,i}$ and $s_{1,1}$.

A given term in Eq.~(\ref{SDU})
arises from a given choice of $(l,j)$: for a term on the first or
second line let us say $l=1$.  Let $(1,1)=F_n(l_0,j_0)$ and
let $(l,j)=F_n(l_0',j_0')$.  
If a matrix is {\it not} trivially moved under on the $n^{\text{th}}$ iteration
then there are two cases:
$(1)$ either $l_0\leq k$ and
$l_0'\leq k$ or $l_0>k$ and $l_0'>k$.  That is, either both matrices
appeared in one of the first $k$ traces, which are traces of products of
conjugates of unitaries, or both matrices appeared in one of the last
$k$ traces, which are traces of unitaries.
Or, case $(2)$: 
$l_0\leq k$ and $l_0'> k$ or
$l_0> k$ and $l_0'\leq k$.  That is, one matrix was in one of the
first $k$ traces and the other was in one of the last $k$ traces.
We then
break the first case into two sub-cases: $(a)$, $j_0=j_0'$ or $(b)$, 
$j_0\neq j_0'$.
We also break the second case into two sub-cases: $(a)$, $j_0=m_1+1-j_0'$
or $(b)$, $j_0\neq m_1+1-j_0'$.
In case $1a$ both matrices are unitary matrices $U(s_{1,j_0})$ or both
are $U(s_{1,j_0})^{\dagger}$ and in case $2a$, one matrix is
$U(s_{1,j_0})$ and the other is $U(s_{1,j_0})^{\dagger}$.
In case $1b$, we know that
$s_{1,j_0}=s_{1,j_0'}$ for $j_0\neq j_0'$
while in case $2b$ we know that $s_{1,j_0}=s_{1,j_0'}+D/2$ for $j_0\neq j_0'$.
Thus, in case $1b$ or $2b$ the term in the Schwinger-Dyson
equation implies some constraint about the choice of
$s_{1,j}$.
To illustrate these different cases, consider the example (\ref{ex1}):
the first term on the right-hand side of the top line is an example
of case $2a$, while the second term on the same line is an example of case $1a$.

Consider a given $j$; if on some iteration and for some $l$
the matrix which was originally
in position $(l,j)$ is not trivially moved and we have case
$1b$ or $2b$, then we can identify some $k$ such that either
$s_{1,j}=s_{1,k}$ or $s_{1,j}=s_{1,k}+D/2$.
Let us write $k=\tau(j)$ in both
cases, for some function $\tau(j)$.
We define a term to have a
``complete rung cancellation of matrix $j$'' if it is not possible
to identify such a $k$ for the given $j$.
We claim that the sum of all terms with a
complete rung cancellation of matrix $i$ is equal to $k!$ so long
as $k\leq N$.
To show this, consider
the product of traces
\be
\label{xsub}
{\rm tr}(U(s_{m}+D/2) ...  U(s_{i+1}+D/2) X^{\dagger} U(s_{i-1}+D/2) ...
U(s_{1}+D/2))^k
{\rm tr}(U(s_{1})...U(s_{i-i}) X U(s_{i+1}) ... U(s_{m})^k,
\ee
where $X$ is some arbitrary unitary matrix.  Averaging
this trace over all unitary matrices $U(s)$ and over all unitary matrices $X$
with the Haar measure,
we find that the trace is equal to $k!$: this can be established by
applying Eq.~(\ref{SDU}) to this trace, and always
cyclically permuting the trace so that $X$ is in the first position.
This calculation is very similar to
the example calculation (\ref{exampleX}) above.
However, applying the Schwinger-Dyson
equations to the trace (\ref{xsub}) without first applying the cyclic
permutation
generates precisely the sum of terms mentioned
above, those in which there is a complete rung cancellation
of matrix $i$.  Thus, this sum of terms equals $k!$.  We further claim that
for any given $i_1,i_2,...,i_d$, the sum of all terms with complete rung
cancellations of matrices $i_1,i_2,...i_d$ is equal to $k!$, as may
be shown by considering a trace in which matrices $U(s_{i_1}),U(s_{i_2}),...$
are replaced by $X_1,X_2,...$, and the trace is averaged over the different
$X_1,X_2,...$.
Then, using the inclusion-exclusion principle,
the sum of terms in which for no $i$ is there a complete rung
cancellation of matrix $i$ is equal to the sum of all terms minus
$k!$.  So, we now focus on the sum of terms with no complete rung cancellations,
which we define to be $E_{0,k}'(s_1,...,s_m)$; if a given choice of
$s_1,...,s_m$ gives rise to a term which terminates at level $n$ with
no complete rung cancellations, then it is possible to identify a
$\tau(i)$ for each $i$.

We now follow the same approach as in \cite{sd} to bound the number
of choices of $s_1,...,s_{m_1^0}$ which can produce a term which terminates
at a level $n$ with no complete rung cancellations.
Given the sequence of choices of terms on the right-hand side of
the Schwinger-Dyson equation (\ref{SDU}), as well as knowledge
of which cancellations occurred at each iteration, we know the
function $\tau(i)$,
and given this function
$\tau(i)$ 
there are now only at most $[D/(D-1)] (D-1)^{m_1^0/2}$
possible values of $s_{1,1},...,s_{1,m_1^0}$.  
Thus, the total number of choices of $s_1,...,s_{m_1^0}$ which can produce
a term which terminates at level $n$ is bounded by
the number of possible choices of terms and
cancellations in the
Schwinger-Dyson equation (\ref{SDU}) at each of the $n$ iterations multiplied
by $[D/(D-1)] (D-1)^{m_1^0/2}$.
At each iteration of the Schwinger-Dyson equations, we make
a particular choice of
$l,j$ at each level, which requires specifying one particular matrix out
of all the matrices on the right-hand side; there are at most $2m_1k-1$ matrices
on the right-hand side, so there are at most $2m_1k-1$
choices (in \cite{sd}, the slightly worse bound $(2m_1k-1)^2$
  was found; we tighten the bound here).
At each such iteration
of the Schwinger-Dyson equations, there may be cancellations in
two different traces if the term came from the second line
of Eq.~(\ref{SDU}), with at most $m_1$ cancellations in each trace, or
cancellations
in two different places of a single trace, if the term came from the fourth
line of Eq.~(\ref{SDU}), with at most $m_1$ cancellations in each place.
Let us call the number
of cancellations $c_1,c_2$ with $0\leq c_1\leq m_1$ and $0\leq c_2 \leq m_1$.
Then,
by specifying $l,j,c_1,c_2$
for each iteration, we succeed in fully specifying how the matrices
move under the $n$ iterations of the Schwinger-Dyson equation; this requires
specifying $n$ numbers ranging from $1...2km_1-1$, and $2n$ numbers
ranging from $0...m_1$.

Thus, there are at most
\be
[D/(D-1)] (D-1)^{m_1^0/2} (2km_1^0-1)^{n} (m_1^0+1)^{2n} \leq [D/(D-1)]
(D-1)^{m_1^0/2} (2km_1^0)^{3n}
\ee
choices of $s_1,...,s_{m_1^0}$ which can produce a term which terminates
at level $n$.
Using Eq.~(\ref{nchoice}),
the number of choices of $s_1,...,s_m$ which can produce a term which
terminates at level $n$ is at most
\be
\label{numnmin}
\sum_{m_1^0=0}^{m}
(D-1)^{m/2} 2^m
(2km_1^0)^{3n}\leq
(D-1)^{m/2} 2^m
\frac{(2km+1)^{3n+1}}{3n+1}.
\ee

For any $s_1,...,s_m$, we define $n_{min}(s_1,...,s_m)$ to be the
smallest level at which a term terminates with no complete rung cancellations.
The sum of terms with $m_1^0=0$, which is the same as the sum
of terms with $n_{min}=0$, is bounded by
\be
\label{ctrepprob}
N^{2k} D^{-m} (D-1)^{m/2} 2^m = N^2 \lambda_H^m.
\ee
Thus, we re-write the sum in Eq.~(\ref{av}) as
\begin{eqnarray}
\label{cutn}
E_{1,k}\leq k!
+N^{2k} N^2 \lambda_H^m
\Bigl( \frac{1}{D} \Bigr)^m \sum_{n=k}^{\infty}
\sum_{s_1=1}^{D}
\sum_{s_2=1}^{D} ...
\sum_{s_m=1}^{D}
\delta_{n_{min}(s_1,...,s_m),n} E_{0,k}'(s_1,...,s_m).
\end{eqnarray}
Therefore, for any
$s_1,...,s_m$ with $n_{min}>0$,
\begin{eqnarray}
\label{e0nmin}
E_{0,k}'(s_1,...,s_m)
&\leq& \sum_{n\geq n_{min}(s_1,...,s_m)} N^{2(k-n)/3}
(2km-1)^n
\\ \nonumber
&=& N^{2k/3} \frac{[N^{-2/3} (2km-1)]^{n_{min}}}{1-N^{-2/3} (2km-1)}.
\end{eqnarray}
From Eqs.~(\ref{numnmin},\ref{cutn},\ref{e0nmin}),
\begin{eqnarray}
E_{1,k}& \leq & k!+\lambda_H^m\Bigl\{ N^2 +N^{2k/3}\sum_{n=k}^{\infty}
\frac{(2km+1)^{3n+1}}{3n+1}
\frac{[N^{-2/3} (2km-1)]^n}{1-N^{-2/3} (2m-1)}\Bigr\}
\\ \nonumber
&\leq &
1+\lambda_H^m\Bigl\{ N^2+
N^{2k/3} \sum_{n=k}^{\infty}
 \frac{2km+1}{(3n+1)[1-N^{-2/3}(2km-1)]}
[N^{-2/3} (2km+1)^4]^n\Bigr\}.
\end{eqnarray}

We then pick $m=(1/4k) N^{1/6}$, so that
$N^{-2/3} (2km+1)^4\leq 1/2$ and
\begin{eqnarray}
|\lambda|& \leq &(E_{1,k}-1)^{1/m} \leq N^{2/m} \lambda_H (1+\cO(1))^{1/m}
\\ \nonumber
&=&\lambda_H (1+\cO(\log(N) k N^{-1/6}).
\end{eqnarray}
Using Markov's inequality,
the probability that $|\lambda|$ is greater than
$c(1+\cO(k\log(N) N^{-1/6})\lambda_{H}(D)$, for any $c\geq 1$, is bounded by
$c^{-(1/4k) N^{1/6}}$.
\qed

\section{Discussion}
We have introduced quantum and classical tensor product expanders.
These provide a way to approximate $t$-designs by acting many times
with a small number of unitaries.  An important open question is whether
efficient implementations of these tensor product expanders exist.

{\it Acknowledgments} AWH thanks Richard Low for catching an error in
the proof of \thm{2c-1q}, as well as useful discussions about
\lemref{mix-projs}. 
MBH thanks the KITP for hospitality while some
of this research was completed.  MBH was supported in part by the
National Science Foundation under Grant No. PHY05-51164 and supported
by U. S. DOE Contract No. DE-AC52-06NA25396.  AWH was supported by the
European Commission under a Marie Curie Fellowship (ASTQIT,
FP-022194), the integrated EC project ``QAP'' (contract
no.~IST-2005-15848), the U.K.~EPSRC, project ``QIP IRC'' and the Army
Research Office under grant W9111NF-05-1-0294.

\appendix
\section{Proof of \lemref{mix-projs}}\label{app:lem-mix}

First, we reduce to the case when the matrices are $2\times 2$ with
$\Pi=\proj{1}$ and $X$ is diagonal.    Express $\|pX + (1-p)Y\|$ as
the maximum of 
$\bra{\psi}pX+(1-p)Y\ket{\psi}$ over all unit vectors $\ket{\psi}$.
Write $\ket{\psi}$ as $\ket{\psi} = \cos(\theta)\ket{\psi_1} +
\sin(\theta)\ket{\psi_2}$, where $0\leq \theta\leq \pi/2$ and
$\ket{\psi_1},\ket{\psi_2}$ are normalized vectors such that
$\Pi\ket{\psi_1}=\ket{\psi_1}$ and
$(I-\Pi)\ket{\psi_2}=\ket{\psi_2}$.  Our conditions on $X$ imply that 
 $\bra{\psi}X\ket{\psi} = \cos^2(\theta) + \bra{\psi_2}X\ket{\psi_2}
 \sin^2(\theta)$ and that $|\bra{\psi_2}X\ket{\psi_2}| \leq 1- \
 \eps_X$.   Next, for $i,j=1,2$ define $Y_{i,j} =
 \bra{\psi_i}Y\ket{\psi_j}$.  Since $\|Y\|\leq 1$, we also have that
 $\| \sum_{i,j=1}^2 Y_{i,j} \ket{i}\bra{j}\|\leq 1$.  We can now replace
 $Y$ with $\sum_{i,j=1}^2 Y_{i,j} \ket{i}\bra{j}$ and $X$ with
 $\proj{1} + \bra{\psi_2}X\ket{\psi_2}\, \proj{2}$.

Now suppose that $|\bra{\psi}X\ket{\psi}| \geq 1 - \eps_X\eps_Y/12$.
Using our bound on $|\bra{\psi_2}X\ket{\psi_2}|$, we obtain
$$ 1 - \frac{\eps_X\eps_Y}{12} \leq 
\cos^2(\theta) + \sin^2(\theta) (1-\eps_X) = 1 - \sin^2(\theta)\eps_X,
$$
implying that $\sin^2(\theta) \leq \eps_Y/12$.  We will show that this
yields an upper bound on $\bra{\psi}Y\ket{\psi}$.

Since $\|Y\|\leq 1$, we have $|Y_{1,2}|, |Y_{2,1}| \leq
\sqrt{1-|Y_{1,1}^2|}$.  
Thus
\ba |\bra{\psi}Y\ket{\psi}| 
& \leq \cos^2(\theta) |Y_{1,1}| + \sin(\theta)\cos(\theta)(|Y_{1,2}| +
|Y_{2,1}|) + \sin^2(\theta) |Y_{2,2}|
\\ & \leq \cos(\theta)|Y_{1,1}| + \sin(\theta)2\sqrt{1 - |Y_{1,1}|^2}
+ \frac{\eps_Y}{12}.\label{eq:theta-times-Y}\ea
If $\theta$ were not constrained then the first two terms of
\eq{theta-times-Y} would be maximized by taking $\theta$ to be
$\hat{\theta}=\arctan(2\sqrt{1-|Y_{1,1}|^2} / |Y_{1,1}|) 
\geq \arctan(2\sqrt{2\eps_Y -\eps_Y^2} / (1-\eps_Y)) \geq
\arctan(2\sqrt{2\eps_Y})$.  Using $\sin^2(\arctan(z)) = z^2 /
(1+z^2),$ we have $\sin^2(\hat{\theta}) \geq 8\eps_Y / (1+8\eps_Y)
\geq  \eps_Y/2$.  Since $\theta$ is constrained to lie in
$[0,\arcsin(\sqrt{\eps_Y/12})]$, it cannot equal $\hat{\theta}$.  Thus
maximizing \eq{theta-times-Y} will require setting $\theta$ to one of
the endpoints of the allowed region.  In particular, the maximum value
of \eq{theta-times-Y} occurs when $\sin^2(\theta)= \eps_Y/12$.  A
similar argument proves that setting $|Y_{1,1}|=1-\eps_Y$ maximizes
\eq{theta-times-Y} as well. Now we calculate
\be |\bra{\psi}Y\ket{\psi}| \leq
(1 - \eps_Y) + 
2\sqrt{\frac{\eps_Y}{12}}\sqrt{2\eps_Y - \eps_Y^2} +
\frac{\eps_Y}{12}
\leq 1 - \l(1-\sqrt{\frac{2}{3}} -\frac{1}{12}\r)\eps_Y
\leq 1 - \frac{\eps_Y}{10}\ee
We have shown that for any $\psi$, either $\bra{\psi}X\ket{\psi}\leq
1-\eps_X\eps_Y/12$ or 
 $\bra{\psi}Y\ket{\psi}\leq
1-\eps_Y/10$.   We now use the triangle inequality to bound
\ban\bra{\psi} p X + (1-p)Y\ket{\psi}  &\leq \max\l( 
p \l(1 - \frac{\eps_X\eps_Y}{12}\r) + (1-p),
p + (1-p)\l(1 - \frac{\eps_Y}{10}\r)\r)
\\ & \leq  1-\frac{\eps_Y}{12}\min(p\eps_X,1-p).\ean
Since this bound applies for all normalized $\ket{\psi}$, it must also
upper-bound $\| pX + (1-p)Y\|$.  Thus  we obtain 
 \eq{intermediate-norm}.  The remaining steps of the Lemma are direct
calculations. \qed

\end{document}